\documentstyle[twocolumn,aps,psfig,prl]{revtex}

\begin{document}
\draft
\preprint{{\bf ETH-TH/98-11}}

\title{\centerline{Thermal Suppression of Strong Pinning}}

\author{Orlando S.\ Wagner, Guido Burkard, Vadim B.\ Geshkenbein, and
  Gianni Blatter} 

\address{Theoretische Physik, ETH-H\"onggerberg, CH-8093 Z\"urich, 
Switzerland} 

\twocolumn[
\date{6 March 1998}
\maketitle
\widetext
\vspace*{-1.0truecm}
\begin{abstract}
\begin{center}
\parbox{14cm}{
  
  We study vortex pinning in layered type-II superconductors in the
  presence of uncorrelated disorder for decoupled layers. Introducing
  the new concept of variable-range thermal smoothing, we describe the
  interplay between strong pinning and thermal fluctuations.  We
  discuss the appearance and analyze the evolution in temperature of
  two distinct non-linear features in the current-voltage
  characteristics. We show how the combination of layering and
  electromagnetic interactions leads to a sharp jump in the critical
  current for the onset of glassy response as a function of
  temperature.}

\end{center}

\end{abstract}
]
\vspace{-0.4truecm}
 
\narrowtext

Quenched disorder in strongly layered superconductors, such as the
Bi-based high-$T_c$ compounds or the organic BEDT-based materials,
naturally leads to the phenomenon of strong vortex pinning.  With the
magnetic field directed perpendicular to the layers the vortex lines
divide up into loosely coupled strings of ``pancake'' vortices
\cite{C}. At low fields, the absence of interactions between the
pancake vortices allows for their free accommodation to the pinning
potential, thus leading to strong pinning of individual pancake
vortices.  This is in contrast to the weak collective pinning
situation \cite{LO}, where the disorder potential competes with
elastic forces, either due to tilt or shear energies.  Whereas a
detailed understanding of the weak-pinning phenomenology has been
developed over the past decade \cite{review}, not much progress has
been made regarding the strong-pinning situation.

\begin{figure} [bt]
\centerline{\psfig{file=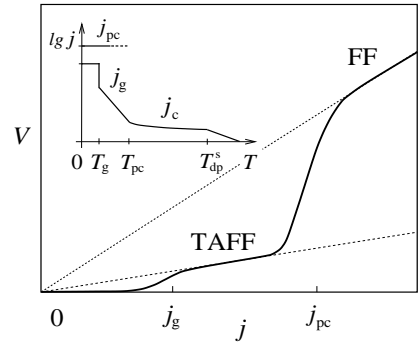,width=5.4cm,height=4.5cm}}
\narrowtext\vspace{2mm}
\caption{\label{V-j-char}Current-voltage characteristics of layered
  superconductors in the pancake-vortex pinning regime ($T < T_{\rm
    pc}$) exhibiting a two-step behavior. Above the critical current
  density $j_{\rm pc}$ we find the usual flux-flow (FF) regime with
  resistivity $\rho_{\rm \scriptscriptstyle FF}$. As the current $j$
  drops below $j_{\rm pc}$, pancake vortices are trapped into
  potential wells of depth $U_{\rm pc}$ and the system enters a second
  ohmic regime with a reduced resistivity due to thermally activated
  flux flow of individual pancake vortices, $\rho_{\rm
    \scriptscriptstyle TAFF} \sim \rho_{\rm \scriptscriptstyle FF}
  \exp (-U_{\rm pc}/T)$. At $j_{\rm g}< j_{\rm pc}$ the motion of
  pancake vortices is inhibited by elastic forces, resulting in a
  sharp drop of the voltage (glassy response).  Inset: temperature
  dependence of the two step features in the CVC. The critical current
  density $j_{\rm pc}$ remains constant up to $T=U_{\rm pc}$, where
  the corresponding step in the CVC disappears. The critical current
  density for glassy response $j_g$ decreases with temperature and, at
  $T=T_{\rm pc}$, smoothly goes over to the critical current density
  $j_c$ determined by weak collective pinning theory. Note that
  $T_{\rm pc} \simeq U_{\rm pc}$ as shown in the text. }
\end{figure}

\begin{figure} [bt]
  \centerline{\psfig{file=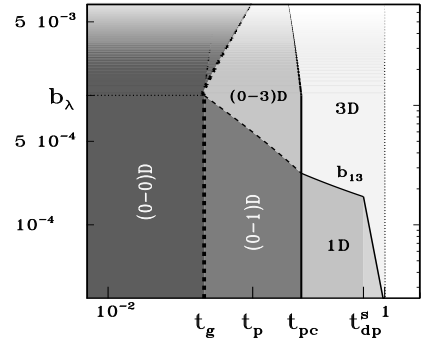,width=5.4cm,height=4.5cm}}
  \narrowtext\vspace{4mm}
\caption{\label{pindiag_em} $B$-$T$-pinning diagram of a layered
  superconductor with decoupled layers (log scales, $t=T/T_c$,
  $b=B/B_{c2}$; $B_\lambda = \Phi_\circ/ \lambda^2$; parameters for
  BiSCCO: $\lambda=1800~{\rm \AA}$, $\xi=25~{\rm \AA}$, $d=15~{\rm
    \AA}$, $T_c = 90~{\rm K}$). Results for the various pinning
  regimes are shown for moderate pinning with $T_p = U_p =10$ K. The
  strong-pinning region at low temperatures $T < T_{\rm pc}$ is
  divided into two parts: At $T<T_g$ thermal effects are irrelevant
  [(0--0)D regime], while at $T_g < T < T_{\rm pc}$ variable-range
  thermal smoothing takes place [(0--1)D and (0--3)D regimes]. The
  dashed fat line at $T=T_g$ indicates the jump in the collective
  pinning length $L_g$ and the critical current density $j_g$. At
  temperatures $T > T_{\rm pc}$ and low inductions vortex segments of
  length $L_c > \lambda$ are pinned collectively (1D regime). The
  remaining area shows the 3D (vortex-bundle) pinning regime.}
\end{figure}

Recent experimental and theoretical interest concentrates on the
low-field properties of vortex matter in strongly layered materials
such as Bi$_2$Sr$_2$Ca$_1$Cu$_2$O$_8$ (BiSCCO) \cite{vB&,Z&,F&,EN,K&},
with particular emphasis on the effects of thermal fluctuations,
quenched disorder, and their mutual interplay.  In this letter, we
present a detailed analysis of the phenomenon of strong vortex pinning
in layered type-II superconductors and its crossover to the
weak-pinning situation due to thermal fluctuations. We assume quenched
point-like disorder and concentrate on the decoupled limit where
electromagnetic interactions between pancake vortices determine the
behavior of vortex matter.  We discuss the presence of two step-like
features in the current-voltage characteristics (CVC) and determine
the evolution in temperature of the two corresponding critical current
densities $j_{\rm pc}(T)$ and $j_g(T)$ describing the onset of strong
pinning and of glassyness, see Fig.~\ref{V-j-char}. To account for
thermal fluctuations smoothing the disorder potential in the
strong-pinning regime we introduce the new concept of {\it
  variable-range} thermal smoothing.  We show how the thermal
depinning of vortices proceeds in a sequence of steps until the usual
weak collective pinning situation is recovered at high temperatures.
Furthermore, we predict that the strong dispersion in the
electromagnetic tilt modulus leads to a pronounced jump in the
critical current density $j_g$ with increasing temperature. The main
results of our analysis are summarized in the pinning diagram of Fig.\ 
\ref{pindiag_em} which shows the various pinning regimes present in
the low-field region below the field $B_\lambda = \Phi_0/\lambda^2$,
where $\lambda$ is the planar London penetration depth and $\Phi_0$
denotes the flux unit. In a forthcoming publication \cite{BWGB} we
will map out the entire pinning diagram and generalize our analysis to
include finite Josephson coupling.
We begin our study with the low-field/low-temperature limit and
consider an individual vortex line oriented perpendicular to the
layers. The presence of point disorder leads to a distortion of the
vortex line with a typical relative displacement $u$ between
neighboring pancake vortices. The optimal pinning state is determined
by the competition between the elastic and the pinning energies.  The
deformation of the vortex line on a length $L$ {\it costs} an energy
${\cal E}_{\rm el}(u,L)\simeq\varepsilon_l (u,k_z\simeq 1/L)\, u^2/L$.
For purely electromagnetic interaction (uncoupled layers) the
elasticity $\varepsilon_l$ takes the strongly dispersive form
$\varepsilon_l (u<\lambda,k_z) = (\varepsilon_\circ/ 2\lambda^2 k_z^2)
\ln [1+\lambda^2 k_z^2/(1+u^2 k_z^2)]$, with the line energy
$\varepsilon_\circ=\left(\Phi_\circ/4\pi\lambda\right)^2$.  Here, we
have interpolated between the formulae valid for $u k_z > 1$ \cite{C}
and $u k_z < 1$ \cite{review}. On the other hand, adjusting to the
disorder potential, a vortex segment of length $L$ {\it gains} the
pinning energy ${\cal E}_{\rm pin}(u,L) \simeq |{\cal E}_0(u)|
\sqrt{L/d}$, where $d$ is the layer separation and ${\cal E}_0(u)$ is
the deepest minimum a pancake vortex can settle in within the area
$u^2$. This energy is determined by the condition $u^2 \int^{{\cal
    E}_0} g({\cal E}) d{\cal E} \simeq 1$, where $g({\cal E})$ is the
distribution of pinning energies, which for a large number of defects
we assume to be Gaussian \cite{K&}
\begin{equation}
g({\cal E})=\frac{1}{\sqrt{\pi} U_p \,\xi^2} \exp\left(-\frac{{\cal
E}^2}{U_p^2}\right).
\label{DOS}
\end{equation}
Here, $U_p$ quantifies the disorder strength and $\xi$ is the planar
coherence length (and also the typical distance between pinstates).
For $u\gg\xi$ (strong pinning) each pancake vortex can explore many
minima and one finds
\begin{equation}
{\cal E}_0(u) \simeq -U_p \left[ \ln\left( \frac{u^2}{\xi^2} \right) 
\right]^{1/2} < -U_p.
\end{equation}
Introducing the energy scale $E_{\rm em} = \varepsilon_\circ d
\xi^2/\lambda^2$, we arrive at the vortex free energy $f$ per unit
length,
\begin{equation}
f(u,L) \simeq \frac{E_{\rm em}}{d} \ln\left(1+\frac{\lambda^2}
{L^2+u^2} \right) \frac{u^2}{\xi^2} + \frac{{\cal E}_0(u)}{d} 
\sqrt{ \frac{d}{L} }.
\label{f_em_u,L}
\end{equation}
Minimizing $f$ with respect to $u$ and $L$ provides us with the
optimal pinning state. For strong pinning the minimum is realized by
the 0D pancake-vortex configuration ($L=d$) and minimizing
Eq.~(\ref{f_em_u,L}) with respect to $u$, we obtain the optimal search
area \cite{K&}
\begin{equation}
u_g^2 \simeq \xi^2 \frac{U_p}{E_{\rm em}} \left[ \ln
\left(\frac{U_p}{E_{\rm em}}\right) \right]^{-1/2}
\left[ \ln\left( \frac{\lambda^2}{\xi^2} \frac{E_{\rm em}}{U_p} 
\right) \right]^{-1}.
\label{ugl}
\end{equation}
The activation barrier for pancake-vortex motion is
\begin{equation}
U_{\rm pc} = - {\cal E}_0 (u_g) \simeq U_p \left[ \ln
\left(\frac{U_p}{E_{\rm em}}\right) \right]^{1/2}.
\label{U_pc}
\end{equation}
Comparing the Lorentz force $j (\Phi_\circ/c) d$ with the pinning
force $U_{\rm pc}/ \xi$, we find the pancake critical current density,
\begin{equation}
j_{\rm pc} \simeq j_p \left[ \ln \left(\frac{U_p}{E_{\rm em}}\right) 
\right]^{1/2}. 
\label{jc_0D}
\end{equation} 
Here, $j_p = j_\circ (U_p/ \varepsilon_\circ d)$ and $j_\circ = c
\varepsilon_\circ / \xi \Phi_\circ$ denotes the depairing current
density.  In the weak-pinning situation $U_p < E_{\rm em}$ we have
$u_g \simeq \xi$, ${\cal E}_0 \simeq -U_p$, and it is energetically
more favorable for the system to settle in the 1D pinning regime:
Minimizing $f(u=\xi,L)$ with respect to $L$, we find $L_c \simeq
\lambda (\lambda/d)^{1/3} (E_{\rm em}/U_p)^{2/3} > \lambda$.  The
parameter $U_p$ can be estimated from experiments measuring the
critical current density at low $B$ and $T$ and is typically of the
order of $10~{\rm K}$ (BiSCCO). In comparison, the electromagnetic
elastic energy $E_{\rm em} \approx 0.2 ~{\rm K}$ is much smaller. From
these estimates we conclude that we usually encounter a strong-pinning
situation with $U_p \gg E_{\rm em}$ in strongly layered high-$T_c$
material.
\vspace{-2mm}
The strong pinning of individual pancake vortices into potential wells
of depth $U_{\rm pc}$ leads to a sharp drop in the current-voltage
characteristics (CVC) at the critical current density $j_{\rm pc}$,
see Fig.~\ref{V-j-char}. However, at finite temperatures the CVC does
not drop to zero, as the individual pancake vortices can overcome
their finite pinning barriers by thermal activation, and we arrive at
a second ohmic regime at small current densities $j<j_{\rm pc}$ with a
reduced resistivity described by thermally activated flux flow
$\rho_{\rm \scriptscriptstyle TAFF} \sim \rho_{\rm \scriptscriptstyle
  FF} \exp (-U_{\rm pc}/T)$ [here, $\rho_{\rm \scriptscriptstyle FF} =
(B/H_{c_2}) \rho_n$ is the usual flux-flow resistivity, with $H_{c_2}$
the upper critical field and $\rho_n$ the normal-state resistivity].
Glassy response appears only at low current densities $j < j_g \ll
j_{\rm pc}$, when the free thermal hopping of pancake vortices is
inhibited by the elastic coupling to other pancake vortices.  To
determine $j_g$ we have to consider the hopping process of individual
pancake vortices. Following the usual variable-range-hopping (VRH)
argument \cite{ES}, a pancake vortex can move freely as long as the
current compensates for the energy $\delta {\cal E}(u)$ required to
hop on to the next favorable state.  The ``minigap'' $\delta {\cal E}
(u)$ is obtained from comparing neighboring favorable states: With
$u^2 \int^{{\cal E}_1} d{\cal E} g({\cal E}) = 2$ we find $\delta
{\cal E}(u) = $ ${\cal E}_1 - {\cal E}_0 \simeq U_p [\ln (u^2/\xi^2)
]^{-1/2}$.  The onset of glassy response is determined by the
condition $j_g (\Phi_\circ/c)$ $ u_g d \simeq \delta {\cal E}(u_g)$,
resulting in the critical current density
\begin{equation}
j_g \simeq j_p \left(\frac{E_{\rm em}}{U_p}\right)^{1/2} \left[ \ln
\left(\frac{U_p}{E_{\rm em}}\right) \right]^{-1/4},
\label{j_g}
\end{equation}
and the corresponding pinning energy
\begin{equation}
U_g = \delta {\cal E}(u_g) \simeq U_p \left[ \ln
\left(\frac{U_p}{E_{\rm em}}\right) \right]^{-1/2}. 
\label{U_g}
\end{equation}
As the external current $j$ decreases below $j_g$, {\it single}
pancake vortices cannot find an appropriate final state any more and
vortex motion involves line segments with a length determined by the
usual laws of creep dynamics \cite{review}, $L(j) \simeq L_g
(j_g/j)^{5/7}$ (here, $L_g=d$).

Going over to finite temperature, we note that the situation remains
unchanged for $T<U_g$ [(0--0)D regime]. As $T$ increases beyond $U_g$,
the process of variable-range thermal smoothing (VRS) sets in: Thermal
fluctuations push the vortices to probe an area $\langle u^2
\rangle_{\rm th} > u_g^2$, but elastic forces prevent individual
pancake vortices from hopping to favorable states, hence segments of
length $L_g(T) > d$ will take over the creep process.  The thermal
hopping of these segments then leads to the smoothing of the pinning
energy. Hence, the temperature $T_g = U_g$ defines a (first) thermal
depinning temperature in our problem.

We proceed with a detailed analysis of VRS: The mean-squared thermal
displacement of a free vortex segment of length $L$ is given by
$\langle u^2(L,T) \rangle_{\rm th} = \int_{1/L}^{1/d} (dk_z/2\pi)
(T/\varepsilon_l(k_z) k_z^2)$, and using the dispersive elasticity
from above, we find
\begin{eqnarray}
\langle u^2(L,T)\rangle_{\rm th}
\simeq \left\{\begin{array}{l r} 
\displaystyle{\xi^2 \frac{T}{E_{\rm em}}}, & d<L<\lambda^2/d,\\
\noalign{\vskip 3 pt}
\displaystyle{\xi^2 \frac{T}{E_{\rm em}} \frac{dL}{\lambda^2}}, 
&\lambda^2/d < L, \end{array} \right.
\label{fluct_em_1D}
\end{eqnarray}
[for $d<L<u$ we should account for a log-correction
$\ln[(\lambda/\xi)^2E_{\rm em}/T]$ guaranteeing the smooth crossover
to $u^2_g$ as $(L,T) \searrow (d,T_g)$]. The thermally smoothed
pinning energy for a vortex segment of length $L$ is given by
\cite{BWGB}
\begin{equation}
{\cal E}_{\rm pin}(L,T) \simeq \sqrt{ \frac{\langle (\Delta 
{\cal E})^2 (T) \rangle}{N(T)}} \sqrt{\frac{L}{d}}, \quad T>T_g.
\label{E_pin_VRS}
\end{equation}
Here, $\langle (\Delta {\cal E})^2 \rangle = \langle {\cal E}^2
\rangle -\langle {\cal E} \rangle^2 $ denotes the fluctuation in the
pinning energy [use Eq.~(\ref{DOS})], which is equal to $U_g^2$ at
$T=T_g$ and tends to $U_p^2/2$ as $T \rightarrow \infty$. $N(T)$ is
the number of available states given the search area $\langle u^2
\rangle_{\rm th}$, $N(T) \approx \langle u^2 \rangle_{\rm th}
\int_{{\cal E}_0} ^{{\cal E}_0+T} d{\cal E} g({\cal E})$, with lower
and upper limits $N(T_g) = 1$ and $N(T \rightarrow \infty) = \langle
u^2 \rangle_{\rm th} /\xi^2$.  It is the suppression in the number of
available states, $N(T) \ll \langle u^2 \rangle_{\rm th} /\xi^2$,
which distinguishes the new VRS from the conventional smoothing
occuring within weak collective pinning theory \cite{FV}. This reduced
smoothing is a consequence of strong pinning and is realized within
the regime $T < T_{\rm pc}$, where the (second) depinning temperature
$T_{\rm pc}$ is determined by the condition $N(T_{\rm pc}) \approx
\langle u^2 \rangle_{\rm th} / 2\xi^2$. For the distribution
(\ref{DOS}) we find $T_{\rm pc} \simeq U_{\rm pc}$. At $T>T_{\rm pc}$
the entire set $N(T) \simeq \langle u^2 \rangle_{\rm th}/ \xi^2$ of
pinning states takes part in the smoothing and Eq.~(\ref{E_pin_VRS})
reproduces the standard expression \cite{FV}, ${\cal E}_{\rm pin}(L,T)
\simeq U_p \sqrt{\xi^2/ \langle u^2 \rangle_{\rm th}} \sqrt{L/d}$.

Within the VRS regime at low temperatures $T_g < T < T_{\rm pc}$ we
use suitable numerical interpolations for the error function [from
$\int^{{\cal E}_0+T}_{{\cal E}_0} d{\cal E} g({\cal E}) {\cal E}^n$,
($n=0,2$)],
\begin{equation}
N(T) \approx \left( \frac{T}{T_g} \right)^{1/2} \!\!\exp 
\left( \frac{T}{T_g} \right),\;\; \langle (\Delta {\cal E})^2 (T)
\rangle \approx T_g T,  
\end{equation}
such that ${\cal E}_{\rm pin}(d,T)$ goes over to $U_g$ and $U_p
\sqrt{\xi^2/ \langle u^2 \rangle_{\rm th}}$ as $T \searrow T_g$ resp.\ 
$T \nearrow T_{\rm pc}$. The vortex free energy $f$ per unit length
takes the form
\begin{eqnarray}
f(L,T) &\simeq& \frac{E_{\rm em}}{d} \ln\left(1+\frac{\lambda^2}
{L^2}\right) \frac{\langle u^2 \rangle_{\rm th}}{\xi^2} \nonumber \\
& & - \frac{T}{d} \left( \frac{T_g}{T} \right)^{3/4} \exp 
\left( -\frac{T}{2 T_g} \right) \sqrt{\frac{d}{L}}.
\label{f_em_L,T}
\end{eqnarray}
Again, the minimum of $f$ with respect to $L$ defines the pinning
length $L_g$. Inspection of Eq.~(\ref{f_em_L,T}) reveals that the
minimum at $L=d$ vanishes at $T=T_g$, implying that $L_g(T)$ is
determined by the minimum at large lengths,
\begin{equation}
L_g(T) \simeq \lambda \left( \frac{\lambda}{d} \right)^{1/3} 
\left(\frac{T}{T_g} \right)^{1/2} \exp \left( \frac{T}{3 T_g} 
\right) < \lambda^2/d.
\end{equation}
The jump from $L_g=d$ to $L_g(T) > \lambda$ at $T=T_g$ implies a
concurring jump in the pinning energy and the critical current
density,
\begin{eqnarray}
U_g(T) &\simeq& T  \left( \frac{\lambda}{d} \right)^{2/3} 
\left( \frac{T_g}{T} \right)^{1/2} \exp \left(- \frac{T}{3 T_g} 
\right),\\
j_g(T) &\simeq& j_g \left( \frac{d}{\lambda} \right)^{2/3} 
\left( \frac{T_g}{T} \right)^{1/2} \exp \left( -\frac{2 T}{3 T_g} 
\right). 
\end{eqnarray}
The jump at $T_g$ is a consequence of the strong dispersion in the
electromagnetic line tension and persists into the high-field regime
\cite{BWGB}, where it matches up with the jump found by Kes and
Koshelev \cite{KK} in their $T=0$ analysis of this problem. A sharp
increase (jump) in the activation energy $U$ by a factor of 10 with
increasing temperature ($T \gtrsim 15$ K) has been found in several
relaxation experiments in BiSCCO material \cite{ZZ}.

At temperatures above $T_{\rm pc}$ we can ignore the underlying strong
pinning. Repeating the above minimization procedure with the usual
smoothed pinning potential \cite{FV}, we obtain the weak-pinning
results (1D regime)
\begin{eqnarray}
L_c(T) &\simeq& \lambda \left(\frac{\lambda}{d}\frac{T^3}{E_{\rm em} 
U_p^2}\right)^{1/3} = \lambda \frac{\lambda}{d}\frac{T}{T^s_{\rm dp}},\\ 
U_c &\simeq& \left(\frac{\lambda^2}{d^2} E_{\rm em}U_p^2\right)^{1/3}
= T_{\rm dp}^s, \label{char_em_1D_it} \\
j_c(T) &\simeq& j_p \left(\frac{d^5}{\lambda^5}\frac{E_{\rm
em}^2}{U_p^2}\right)^{1/3} \left(\frac{T_{\rm dp}^s}{T}\right)^{3/2}, 
\end{eqnarray}
where we have introduced the single-vortex depinning temperature
$T_{\rm dp}^s$. Quick inspection shows that $L_c (T)$, $U_c$, and $j_c
(T)$ match up with $L_g (T)$, $U_g(T)$, and the onset of glassy
response at $j_g(T)$ when $T \searrow T_{\rm pc}$.

When the collective pinning length $L_c(T)$ exceeds $\lambda^2/d$ the
mean thermal displacement grows with increasing length and pinning
becomes marginal \cite{review}.  The determination of $L_c(T)$
involves the calculation of the disorder-induced fluctuations
$\langle\langle u_p^2(L,T) \rangle\rangle$, where $\langle\langle
\cdot \rangle\rangle$ denotes averaging over thermal fluctuations and
disorder.  Repeating this calculation for the case of electromagnetic
coupling, we find $\langle\langle u_p^2(L,T) \rangle\rangle/\xi^2
\simeq$ $(U_p/T)^2 (L/d) \ln(Ld/\lambda^2)$. As the disorder-induced
fluctuations increase beyond the ther\-mal ones, the system crosses
over to the pinning-domi\-nated regime. The condition $\langle\langle
u_p^2(L,T) \rangle\rangle \simeq \langle u^2(L,T) \rangle_{\rm th}$
determines the length
\begin{equation}
L_c(T) \simeq \lambda \frac{\lambda}{d} 
\exp\left(\frac{T}{T_{\rm dp}^s}\right)^3,
\label{Lc_em_1D_ht}
\end{equation}
and the critical current density takes the form
\begin{equation}
j_c(T) \simeq j_p \!\left(\frac{d^5}{\lambda^5} \frac{E_{\rm
em}^2}{U_p^2}\right)^{\frac{1}{3}} \!\!\sqrt{\frac{T}{T^s_{\rm
dp}}} \exp\left[ -\frac{3}{2}\left(\frac{T}{T_{\rm
dp}^s}\right)^3 \right]. 
\label{j_c_T_m}
\end{equation}
The above results properly match the previous ones at the crossover
temperature $T = T_{\rm dp}^s$. The complete temperature dependencies
of the (critical) current densities $j_{\rm pc}$, $j_c$, and $j_g$ are
illustrated in the inset of Fig.~\ref{V-j-char}.

As we increase the magnetic field, interactions with other vortices
start to interfere with the accommodation of the individual vortices
to the disorder potential. The regime of validity of the above results
is found by comparing the tilt elastic energy $E_{\rm em} (u/\xi)^2$
with that of shear, $c_{\scriptscriptstyle 66} u^2 d$, with the shear
modulus given by $c_{\scriptscriptstyle 66} \simeq \varepsilon_\circ/
a_\circ^2$ and $(\varepsilon_\circ / \lambda^2) \sqrt{\lambda/
  a_\circ} \exp( -a_\circ/ \lambda )$ above resp.\ below $B_\lambda$.
At low temperatures $T<T_g$ we find that pancake vortices start to
interact within the planes as the magnetic induction increases beyond
$B_\lambda$, see Fig.~\ref{pindiag_em}. At $T>T_g$ we have to account
for the finite lengths $L_g(T)$ and $L_c(T)$ of the segments involved.
The crossover condition $L \simeq \lambda \exp(a_\circ/2\lambda)$
produces the result
\begin{eqnarray}
B_{\scriptscriptstyle 13}(T) &\simeq& B_\lambda \left(\frac{T}{T_g} 
+\ln \frac{\lambda}{d}\right)^{-2},\quad T_g < T < T_{\rm pc},\\
B_{\scriptscriptstyle 13} (T) &\simeq& B_\lambda \ln^{-2} \left(
\frac{\lambda}{d} \frac{T^3}{E_{\rm em}U_p^2} \right), 
\quad T_{\rm pc} < T < T_{\rm dp}^s,\\ 
B_{\scriptscriptstyle 13} (T) &\simeq& B_\lambda
\left( \ln\frac{\lambda^2}{d^2} \right)^{-2} \! \left(\frac{T_{\rm
dp}^s}{T}\right)^6, \quad T_{\rm dp}^s < T.
\end{eqnarray}
Above $B_{\scriptscriptstyle 13} (T)$ pinning involves relaxation of
vortex bundles [(0-3)D and 3D regimes]. These results apply to the
vortex solid regime below the melting line $B_{\rm m}(T)$ \cite{BGLN}:
Upon melting, both the shear and tilt moduli vanish in the resulting
pancake-vortex gas phase, cutting off the 1D and 3D pinning regimes at
$B_{\rm m}(T)$.

Our analysis (and its generalization to higher fields \cite{BWGB})
sheds light on two recent experiments in layered BiSCCO material. The
Maley analysis of the creep barrier as measured by van der Beek {\it
  et al.}  \cite{vB&} is consistent with a diverging barrier
$U(j\rightarrow 0) \rightarrow \infty$, rather than the constant
barrier expected for strong pancake pinning in the low-$T$/low-$B$
domain.  The above results explain how individual pancake vortices
couple into vortex lines exhibiting diverging barriers as the driving
force vanishes. Second, recent local Hall-probe measurements of the
current flow in BiSCCO crystals show a sharp, roughly
field-independent onset of bulk pinning as the temperature decreases
below $T \approx 40$ K \cite{F&}. This experimental finding is in
agreement with the appearance of strong non-linearities in the CVC
below the temperature $T_{\rm pc}$, see Fig.~\ref{V-j-char}.  The
definite identification of this pinning onset with either $j_{\rm pc}$
or $j_g$ requires a detailed amplitude and frequency analysis of the
experimental feature.

We thank A.E.\ Koshelev, M.\ Nider\"ost, and A.\ Suter for stimulating
discussions, and the Swiss National Foundation for financial support.


\vspace{-0.4truecm}

\begin{thebibliography}{99}
  
\vspace{-1.4truecm}

\bibitem{C} J.R.\ Clem, Phys.\ Rev.\ B {\bf 43}, 7837 (1991).
  
\bibitem{LO} A.I.\ Larkin and Yu.N.\ Ovchinnikov, J.\ Low Temp.\ Phys.
  {\bf 34}, 409 (1979).
  
\bibitem{review} G.\ Blatter {\it et al.}, Rev.\ Mod.\ Phys.\ {\bf
    66}, 1125 (1994).
  
\bibitem{vB&} C.J.\ van der Beek {\it et al.}, Physica C {\bf 192},
  307 (1992).
  
\bibitem{Z&} E.\ Zeldov {\em et al.}, Nature (London) {\bf 375}, 373
  (1995) and Europhysics Lett. {\bf 30}, 367 (1995).
  
\bibitem{F&} D.T.\ Fuchs {\it et al.}, unpublished.
  
\bibitem{EN} D.\ Erta\c{s} and D.R.\ Nelson, Physica C {\bf 272}, 79
  (1996); T.\ Giamarchi and P.\ Le Doussal, Phys.\ Rev.\ B {\bf 55},
  6577 (1997).
  
\bibitem{K&} A.E.\ Koshelev {\it et al.}, Phys.\ Rev.\ B {\bf 53},
  2786 (1996); A.E.\ Koshelev and V.M.\ Vinokur, preprint cond-mat/
  9801144.
  
\bibitem{BWGB} G.\ Burkard, O.S.\ Wagner, V.B.\ Geshkenbein, and G.\ 
  Blatter, unpublished.
 
\bibitem{ES} A.L.\ Efros and B.I.\ Shklovskii, {\it Electronic
    Properties of Doped Semiconductors}, (Springer, Berlin, 1984).
  
\bibitem{FV} M.V.\ Feigel'man and V.M.\ Vinokur, Phys.\ Rev.\ B {\bf
    41}, 8986 (1990).
  
\bibitem{KK} A.E.\ Koshelev and P.\ Kes, Phys.\ Rev.\ B {\bf 48}, 6539
  (1993).
  
\bibitem{ZZ} V.N.\ Zavaritzky and N.V.\ Zavaritzky, Physica C {\bf 185
    -- 189}, 2141 (1991); V.V.\ Metlushko {\it et al.}, Physica B {\bf
    194 -- 196} 2219 (1994); M.\ Nider\"ost {\it et al.}, Phys.\ Rev.\ 
  B {\bf 53}, 9286 (1996).
  
\bibitem{BGLN} G.\ Blatter {\it et al.}, Phys.\ Rev.\ B {\bf 54}, 72
  (1996).

\end{thebibliography}
\end{document}